# Thermal radiation in systems of many dipoles

Eric Tervo,[1,*] Mathieu Francoeur,[2] Baratunde Cola,[1,3] and Zhuomin Zhang[1,†]

[1]*George W. Woodruff School of Mechanical Engineering, Georgia Institute of Technology, Atlanta, GA*
[2]*Department of Mechanical Engineering, University of Utah, Salt Lake City, UT*
[3]*School of Materials Science and Engineering, Georgia Institute of Technology, Atlanta, GA*
*eric.tervo@gatech.edu
†zhuomin.zhang@me.gatech.edu

Systems of many nanoparticles or volume-discretized bodies exhibit collective radiative properties that could be used for enhanced, guided, or tunable thermal radiation. These are commonly treated as assemblies of point dipoles with interactions described by Maxwell's equations and thermal fluctuations correlated by the fluctuation-dissipation theorem. Here, we unify different theoretical descriptions of these systems and provide a complete derivation of many-dipole thermal radiation, showing that the correct use of the fluctuation-dissipation theorem depends on the definitions of fluctuating and induced dipole moments. We formulate a method to calculate the diffusive radiative thermal conductivity of arbitrary collections of nanoparticles; this allows the comparison of thermal radiation to other heat transfer modes and across different material systems. We calculate the radiative thermal conductivity of ordered and disordered arrays of SiC and $SiO_2$ nanoparticles and show that thermal radiation can significantly contribute to thermal transport in these systems. We validate our calculations by comparison to the exact solution for a one-dimensional particle chain, and we demonstrate that the dipolar approximation significantly underpredicts the exact results at separation distances less than the particle radius.

## I. Introduction

When light strikes a collection of small, closely spaced particles, the energy absorbed and scattered depends on electromagnetic interactions among all particles in the group [1,2]. Similarly, thermal radiation in a system of many nanostructures is governed by collective behavior arising from many-body interactions [3]. While classical light scattering from small particles has a long history of study and is relatively well-understood, thermal radiation in such systems has only been explored since the early 2000s [4,5], particularly for objects which support resonant surface modes (surface polaritons). In the last decade, significant progress has been made on this front with the development of kinetic theories for ordered arrays of plasmonic resonators [4-9], exact many-body theories for collections of small particles [3,10-19], and the thermal discrete dipole approximation (T-DDA) for arbitrary bodies [14,20-22] analogous to the discrete dipole approximation for electromagnetic scattering calculations [23].

Research on thermal radiation between nanostructures has been motivated by an improved understanding of near-field effects, which can cause the energy exchange at sub-wavelength separation distances to significantly exceed the far-field blackbody limit [24]. This enhancement of energy transfer was first formulated by Polder and Van Hove based on the fluctuation-dissipation theorem in 1971 [25], has been demonstrated by several groups in the past decade [26], and has applications in thermal management [27], energy conversion [28,29], and scanning thermal microscopy [30], among others. Although most studies have focused on radiation between two objects or one-dimensional layered structures, a growing body of work focuses on many-body interactions. These have been used to predict interesting and potentially useful effects such as many-body amplification of thermal radiation [3], heat superdiffusion in systems of plasmonic particles [31], spectral redshifts in near-field thermal spectroscopy [32], and several magneto-optical phenomena [33-36] such as a thermal Hall effect [33].

Many-body radiative heat transfer theory integrates fluctuating thermal currents into Maxwell's equations to describe electromagnetic interactions between particles modeled as point dipoles. The foundations of this theory were introduced in the early 2010s for spherical nanoparticles by Ben-Abdallah *et al.* [3] and Messina *et al.* [10], and subsequent research has adapted the theory for anisotropic particles [12], magnetic polarization [14,15], core-shell particles [17], and point multipoles [18]. A similar approach has been developed by Edalatpour and Francoeur [20-22] for a numerically exact method of describing radiative transfer between arbitrary volume-discretized bodies, known as the T-DDA. However, a key difference exists between the many-body theory and the T-DDA: the many-body method uses an *exciting* field formalism, where the induced dipole moments are related to the incident external electric fields, while the T-DDA uses an *actual* field formalism, where the induced dipole moments are related to the total macroscopic electric fields. This in turn requires different forms of the fluctuation-dissipation theorem, which describes the fluctuating thermal currents [37-41]. The distinction between the two approaches is similar to that between two numerical methods for classical electromagnetic scattering, the coupled dipole method (*exciting* field formalism) and the method of moments (*actual* field formalism), which have been shown to be mathematically equivalent [42].

In this work, we demonstrate that the many-body method and T-DDA are also mathematically equivalent, unifying the two theoretical approaches and clarifying the appropriate use of the fluctuation-dissipation theorem. We show that when an actual field formalism is used (T-DDA approach), the fluctuation-dissipation theorem may be used directly, in contrast to an exciting field formalism (many-body approach) that requires the correlation function to be determined from the emitted thermal fields [10]. We also describe a computational method to calculate the effective diffusive radiative thermal conductivity of large arbitrary systems of dipoles. The many-body and T-DDA methods allow the calculation of radiation exchange between pairs of dipoles or total power absorbed by individual dipoles. However, in some circumstances, these quantities may not be very useful by themselves, such as when comparing modes of heat transfer in nanoparticle beds or arrays of plasmonic resonators [8,9,31,43]. In these situations, an effective thermal conductivity due to thermal radiation permits direct comparison against phonon thermal conduction in addition to providing a more familiar figure of merit for thermal applications. We apply our method to ordered and disordered arrays of SiC and $SiO_2$ nanoparticles, and we validate the method against the exact solution for linear chains of spherical particles [19].

## II. Theory

### A. Fluctuating and Induced Dipole Moments

We consider a system of $N$ nonmagnetic nanoparticles or subvolumes modeled as electric point dipoles immersed in a transparent background medium of relative permittivity $\varepsilon_m$. The total electric field at location $\mathbf{r}$ and frequency $\omega$ is then given by both the many-body method [10] and the T-DDA [21] as

$$\mathbf{E}(\mathbf{r},\omega) = \frac{k^2}{\varepsilon_m \varepsilon_0} \sum_i \mathbb{G}^{(0)}(\mathbf{r},\mathbf{r}_i,\omega) \cdot \mathbf{p}(\mathbf{r}_i,\omega) \tag{1}$$

where $k^2 = \varepsilon_m \omega^2/c^2$ with $c$ the speed of light in vacuum, $\varepsilon_0$ is the vacuum permittivity, $\mathbf{p}(\mathbf{r}_i,\omega)$ is the total dipole moment of the $i$th dipole, and the summation runs over all dipoles. $\mathbb{G}^{(0)}(\mathbf{r},\mathbf{r}_i,\omega)$ is the dyadic Green's function for the background medium given by

$$\mathbb{G}^{(0)}(\mathbf{r},\mathbf{r}_i,\omega) = \frac{\mathrm{e}^{ikR}}{4\pi R}\left[\left(1+\frac{ikR-1}{k^2R^2}\right)\mathbb{I} + \frac{3(1-ikR)-k^2R^2}{k^2R^2}\hat{\mathbf{R}}\otimes\hat{\mathbf{R}}\right] \tag{2}$$

where $\mathbb{I}$ is the identity matrix, $\mathbf{R} = \mathbf{r} - \mathbf{r}_i$, $R = |\mathbf{R}|$, and $\hat{\mathbf{R}} = \mathbf{R}/R$. Here we have neglected the field arising from an external bath as considered in reference [10] for simplicity and correspondence with the T-DDA [21]. Modeling a particle or subvolume as a point dipole requires that the electric field and dyadic Green's function are constant throughout the volume of each object, which in practice means that its size must be much smaller than the wavelengths and decay lengths of the electric fields under consideration [21]. The total dipole moment is split into a fluctuating part due to fluctuating thermal currents and an induced part due to electric fields:

$$\mathbf{p}_i = \mathbf{p}_i^{(\text{fl})} + \mathbf{p}_i^{(\text{ind})} \tag{3}$$

where $\mathbf{p}(\mathbf{r}_i, \omega)$ is written as $\mathbf{p}_i$ for compactness.

At this point the many-body and T-DDA methods diverge, using different definitions for the fluctuating and induced dipole moments. The many-body method expresses $\mathbf{p}_i^{(\text{ind,MB})} = \varepsilon_0 \alpha_i^{(\text{d})} \mathbf{E}_i^{(\text{exc})}$ with $\mathbf{E}_i^{(\text{exc})}$ described by Equation (1) in terms of the *exciting* field or the fields from all other dipoles as [10]

$$\mathbf{p}_i^{(\text{ind,MB})} = \alpha_i^{(\text{d})} \frac{k^2}{\varepsilon_m} \sum_{j \neq i} \mathbb{G}_{ij}^{(0)} \cdot \mathbf{p}_j \tag{4}$$

where the summation runs over all dipoles *except* the $i$th dipole, and $\alpha_i^{(\text{d})}$ is the dressed polarizability, which accounts for radiation damping and is written as

$$\alpha_i^{(\text{d})} = \left( \frac{1}{\alpha_i^{(\text{CM})}} - \frac{ik^3}{6\pi\varepsilon_m} \right)^{-1} \tag{5}$$

Here, $\alpha_i^{(\text{CM})}$ is the Clausius-Mossotti polarizability defined as

$$\alpha_i^{(\text{CM})} = 3\varepsilon_m V_i \frac{\varepsilon_i - \varepsilon_m}{\varepsilon_i + 2\varepsilon_m} \tag{6}$$

where $V_i$ is the volume of the $i$th particle or subvolume and $\varepsilon_i$ is its complex relative permittivity. Equation (5) is specific to spherical or cubical volumes, and Equation (6) is for spherical volumes, but these may be modified for other shapes [1,44]. The definition in Equation (4) is consistent with the coupled dipole method commonly used in the discrete dipole approximation for electromagnetic scattering calculations [23,42]. It is important to note, however, that $\mathbf{p}_i^{(\text{ind,MB})}$ does not include the parts of the dipole moment due to its own thermal fluctuations nor due to the fields produced by its own thermal fluctuations. These parts must, therefore, both be included in $\mathbf{p}_i^{(\text{fl})}$ in order to obtain the correct total dipole moment $\mathbf{p}_i$.

The T-DDA on the other hand expresses $\mathbf{p}_i^{(\text{ind,TDDA})} = \varepsilon_0 \alpha_i^{(0)} \mathbf{E}_i^{(\text{act})}$ with $\mathbf{E}_i^{(\text{act})}$ described by Equation (1) in terms of the *actual* total field at the $i$th dipole as [21]

$$\mathbf{p}_i^{(\text{ind,TDDA})} = \alpha_i^{(0)} \frac{k^2}{\varepsilon_m} \sum_{j} \mathbb{G}_{ij}^{(0)} \cdot \mathbf{p}_j \tag{7}$$

where the summation runs over all dipoles *including* the $i$th dipole, and $\alpha_i^{(0)}$ is the bare polarizability defined as

$$\alpha_i^{(0)} = V_i (\varepsilon_i - \varepsilon_m) \tag{8}$$

A derivation of Equation (7) from Maxwell's equations is provided in Appendix A for completeness. This definition is consistent with the method of moments [42]. Since $\mathbf{p}_i^{(\mathrm{ind,TDDA})}$ includes the effects of all fields on the dipole moment, including those produced by its own fluctuating currents, the $\mathbf{p}_i^{(\mathrm{fl})}$ term must only include the part directly due to fluctuating currents in order to obtain the correct total dipole moment $\mathbf{p}_i$. As we shall see, this allows a more straightforward use of the fluctuation-dissipation theorem. The remainder of this derivation will use the T-DDA's definition of the induced dipole moment, but we will still recover the same final equation for heat transfer derived via the many-body method.

## B. Expressions for Total Dipole Moments and Electric Fields

Combining Equations (3) and (7) yields a set of $N$ self-consistent equations:

$$\mathbf{p}_i = \mathbf{p}_i^{(\mathrm{fl})} + \alpha_i^{(0)} \frac{k^2}{\varepsilon_m} \sum_j \mathbb{G}_{ij}^{(0)} \cdot \mathbf{p}_j \tag{9}$$

When $j = i$, the dyadic Green's function has a singularity, so this value must be treated separately. $\mathbb{G}_{ii}^{(0)}$ may be evaluated for $V_i$ using the principal value method, assuming $V_i$ is spherical or may approximated by a sphere of equivalent volume [21,44,45]. For a spherical or cubical volume, this results in

$$\mathbb{G}_{ii}^{(0)} = \left(-\frac{1}{3V_i k^2} + \frac{ik}{6\pi}\right)\mathbb{I} \tag{10}$$

Separating the $\mathbb{G}_{ii}^{(0)}$ term from the summation in Equation (9) and rearranging yields

$$\left(\frac{1}{\alpha_i^{(0)}} - \frac{k^2}{\varepsilon_m}\mathbb{G}_{ii}^{(0)}\right)\mathbf{p}_i = \frac{1}{\alpha_i^{(0)}}\mathbf{p}_i^{(\mathrm{fl})} + \frac{k^2}{\varepsilon_m}\sum_{j\neq i}\mathbb{G}_{ij}^{(0)} \cdot \mathbf{p}_j \tag{11}$$

The term in parenthesis on the left-hand side is identical to the inverse of the dressed polarizability (given by Equations (5) and (6) for spherical volumes) [11]:

$$\frac{1}{\alpha_i^{(\mathrm{d})}} = \frac{1}{\alpha_i^{(0)}} - \frac{k^2}{\varepsilon_m}\mathbb{G}_{ii}^{(0)} \tag{12}$$

Equation (11) is the same as the main equation derived in the T-DDA [21], except with a different form of the dressed polarizability due to a different treatment of $\mathbb{G}_{ii}^{(0)}$.

We can now follow the approach of the many-body method [10] to find explicit expressions for $\mathbf{p}_i$ and $\mathbf{E}_i$ in terms of the fluctuating parts of the dipole moments. Rearranging Equation (11) with the use of Equation (12) and writing it in matrix form gives us the relation:

$$\begin{pmatrix} \mathbf{p}_1 \\ \vdots \\ \mathbf{p}_N \end{pmatrix} = \mathbb{A}^{-1} \begin{pmatrix} \dfrac{\alpha_1^{(\mathrm{d})}}{\alpha_1^{(0)}}\mathbf{p}_1^{(\mathrm{fl})} \\ \vdots \\ \dfrac{\alpha_N^{(\mathrm{d})}}{\alpha_N^{(0)}}\mathbf{p}_N^{(\mathrm{fl})} \end{pmatrix} \tag{13}$$

This equation is similar to that derived in reference [10] with the same $3N \times 3N$ interaction matrix, but the vector of fluctuating thermal currents has additional $\alpha_j^{(\mathrm{d})}/\alpha_j^{(0)}$ coefficients here. $\mathbb{A}$ is defined by its $N^2$ $3 \times 3$ submatrices with $i, j = 1 \ldots N$:

$$\mathbb{A}_{ij} = \delta_{ij}\mathbb{I} - \left(1 - \delta_{ij}\right)\alpha_i^{(\mathrm{d})}\frac{k^2}{\varepsilon_m}\mathbb{G}_{ij}^{(0)} \tag{14}$$

where $\delta_{ij}$ is the Kronecker delta function. Inserting Equation (13) into Equation (1) gives us an expression for the local fields:

$$\begin{pmatrix}\mathbf{E}_1\\ \vdots \\ \mathbf{E}_N\end{pmatrix} = \mathbb{D}\mathbb{A}^{-1}\begin{pmatrix}\dfrac{\alpha_1^{(\mathrm{d})}}{\alpha_1^{(0)}}\mathbf{p}_1^{(\mathrm{fl})}\\ \vdots \\ \dfrac{\alpha_N^{(\mathrm{d})}}{\alpha_N^{(0)}}\mathbf{p}_N^{(\mathrm{fl})}\end{pmatrix} \tag{15}$$

which once again is similar to that derived in reference [10] except for the coefficients on $\mathbf{p}_j^{(\mathrm{fl})}$. $\mathbb{D}$ is defined by its submatrices as

$$\mathbb{D}_{ij} = \frac{k^2}{\varepsilon_m\varepsilon_0}\mathbb{G}_{ij}^{(0)} \tag{16}$$

As will be seen later, it is helpful to express $\mathbb{D}\mathbb{A}^{-1}$ in the same form as reference [10]:

$$\mathbb{D}\mathbb{A}^{-1} = (\mathbb{B}\mathbb{A}^{-1} - \mathbb{C}^{-1}), \qquad \mathbb{B}_{ij} = \delta_{ij}\left(\frac{1}{\varepsilon_0\alpha_i^{(\mathrm{d})}}\mathbb{I} + \frac{k^2}{\varepsilon_m\varepsilon_0}\mathbb{G}_{ii}^{(0)}\right), \qquad \mathbb{C}_{ij} = \delta_{ij}\varepsilon_0\alpha_i^{(\mathrm{d})}\mathbb{I} \tag{17}$$

## C. Energy Exchange and the Fluctuation-Dissipation Theorem

With the expressions for total dipole moment and local electric field in hand, we can now turn our attention to the energy exchange. Using the convention for the Fourier transform $f(t) = \int \frac{d\omega}{2\pi} f(\omega)\mathrm{e}^{-i\omega t}$, the power absorbed by the $i$th dipole in the time domain is [10,14]

$$Q_i(t, T_1, \ldots, T_N) = 2\int_0^\infty \frac{d\omega}{2\pi}\omega\int_0^\infty \frac{d\omega'}{2\pi}\mathrm{Im}\left[\langle \mathbf{p}_i(\omega)\cdot\mathbf{E}_i^*(\omega')\rangle \mathrm{e}^{-i(\omega-\omega')t}\right] \tag{18}$$

where $T_i$ is the temperature of the $i$th dipole, and the angled brackets represent the ensemble average. Focusing on this correlation function and inserting Equations (13) and (15) yields

$$\langle \mathbf{p}_i(\omega)\cdot\mathbf{E}_i^*(\omega')\rangle = \sum_\beta\sum_{jj'}\sum_{\gamma\gamma'}\left[(A^{-1})_{ij,\beta\gamma}(BA^{-1} - C^{-1})'^*_{ij',\beta\gamma'}\frac{\alpha_j^{(\mathrm{d})}\alpha_{j'}'^{(\mathrm{d})*}}{\alpha_j^{(0)}\alpha_{j'}'^{(0)*}}\langle p_{j,\gamma}^{(\mathrm{fl})}p_{j',\gamma'}'^{(\mathrm{fl})*}\rangle\right] \tag{19}$$

Here the primes indicate that the variables are a function of $\omega'$, and the Greek subscripts are indices 1, 2, 3 representing the Cartesian components of vector quantities. For example, $(A^{-1})_{ij,\beta\gamma}$ is the $[\beta,\gamma]$ element of the $3\times 3$ matrix $\mathbb{A}_{ij}^{-1}$. The correlation function $\langle p_{j,\gamma}^{(\mathrm{fl})}p_{j',\gamma'}'^{(\mathrm{fl})*}\rangle$ is provided by the fluctuation-dissipation theorem [37-41]. For a dipole moment with a polarizability defined by $\mathbf{p}_j = \varepsilon_0\alpha_j\mathbf{E}_j$, we have [40,41]

$$\langle p_{j,\gamma}^{(\mathrm{fl})}(\omega)p_{j',\gamma'}^{(\mathrm{fl})*}(\omega')\rangle = 4\pi\hbar\varepsilon_0 n\left(\omega, T_j\right)\mathrm{Im}\left(\alpha_j\right)\delta_{jj'}\delta_{\gamma\gamma'}\delta(\omega-\omega') \tag{20}$$

where $n(\omega, T_j)$ is the Bose-Einstein distribution function. However, care must be taken to use this relation appropriately. In the present case, the bare polarizability $\alpha_j^{(0)}$ satisfies $\mathbf{p}_j = \varepsilon_0 \alpha_j \mathbf{E}_j$ for $\alpha_j$, and $\mathbf{p}_j^{(\mathrm{fl})}$ was defined to only include the part of the dipole moment due to thermal fluctuations. This means $\alpha_j^{(0)}$ may be used directly for the polarizability in the fluctuation-dissipation theorem. Equation (20) used with $\alpha_j^{(0)}$ may also be derived directly from the fluctuation-dissipation theorem for thermal currents as done in reference [21] and shown in Appendix B. In the many-body theory [10], $\mathbf{p}_j^{(\mathrm{fl})}$ was defined to include the parts of the dipole moment due to fluctuating thermal currents as well as the fields produced by those fluctuations, as described in Section II.A. Furthermore, the many-body derivation uses only the dressed polarizability, which does not satisfy $\mathbf{p}_j = \varepsilon_0 \alpha_j \mathbf{E}_j$ because $\mathbf{E}_j$ is the actual field instead of the exciting field. For these reasons, the appropriate fluctuation-dissipation theorem in the many-body approach must instead be derived by considering the fields emitted by a single dipole in equilibrium with a thermal bath, as described by Messina *et al.* [10] and Sääskilahti *et al.* [11]. The resulting fluctuation-dissipation theorem contains a reduced absorption factor $\chi_j = \mathrm{Im}\left(\alpha_j^{(\mathrm{d})}\right) - \frac{k^3}{6\pi\varepsilon_m}\left|\alpha_j^{(\mathrm{d})}\right|^2$ used in place of $\mathrm{Im}(\alpha_j)$ in Equation (20). This factor was first derived by Manjavacas and de Abajo [41,46]. We stress that this factor is not inherent to the fluctuation-dissipation theorem but arises due to the definition of the fluctuating and induced dipole moments. That the bare polarizability may instead be used directly with the same results (if $\mathbf{p}_j^{(\mathrm{ind})}$ and $\mathbf{p}_j^{(\mathrm{fl})}$ are defined appropriately) is a key result of this paper.

Using the fluctuation-dissipation theorem with $\alpha_j^{(0)}$ as the polarizability, applying the definitions of $\mathbb{B}$ and $\mathbb{C}$ from Equation (17), and reverting to matrix notation, we can rewrite Equation (19) as

$$\begin{aligned}\langle \mathbf{p}_i(\omega)\cdot\mathbf{E}_i^*(\omega')\rangle \\ &= 4\pi\hbar\varepsilon_0\delta(\omega-\omega') \\ &\times\left[\sum_j \frac{\left|\alpha_j^{(\mathrm{d})}\right|^2 \mathrm{Im}\left(\alpha_j^{(0)}\right)}{\left|\alpha_j^{(0)}\right|^2} n(\omega,T_j)\left(\frac{1}{\varepsilon_0\alpha_i^{(\mathrm{d})}} + \frac{k^2}{\varepsilon_m\varepsilon_0}\mathbb{G}_{ii}^{(0)}\right)^* \mathrm{Tr}\left(\mathbb{A}_{ij}^{-1}\mathbb{A}_{ij}^{-1\dagger}\right)\right. \\ &\left.- \frac{\left|\alpha_i^{(\mathrm{d})}\right|^2 \mathrm{Im}\left(\alpha_i^{(0)}\right)}{\left|\alpha_i^{(0)}\right|^2} n(\omega,T_i)\frac{1}{\varepsilon_0\alpha_i^{(\mathrm{d})*}}\mathrm{Tr}\left(\mathbb{A}_{ii}^{-1}\right)\right]\end{aligned} \tag{21}$$

where we have used the $\delta(\omega-\omega')$ function to eliminate the dependence on $\omega'$ but still retained it here as it affects other parts of Equation (18). We are only concerned with the imaginary part as shown in Equation (18). The integrand can be simplified by using the definition of dressed polarizability from Equation (12) and becomes

$$\begin{aligned}\mathrm{Im}\big[\langle \mathbf{p}_i(\omega)\cdot \mathbf{E}_i^*(\omega')\rangle e^{-i(\omega-\omega')t}\big] &= 4\pi\hbar\delta(\omega-\omega') \\ &\times\left[\sum_j \frac{\left|\alpha_j^{(\mathrm{d})}\right|^2 \mathrm{Im}\left(\alpha_j^{(0)}\right)}{\left|\alpha_j^{(0)}\right|^2}\frac{\mathrm{Im}\left(\alpha_i^{(0)}\right)}{\left|\alpha_i^{(0)}\right|^2} n(\omega,T_j)\mathrm{Tr}\left(\mathbb{A}_{ij}^{-1}\mathbb{A}_{ij}^{-1\dagger}\right)\right. \\ &\left. - \frac{\left|\alpha_i^{(\mathrm{d})}\right|^2 \mathrm{Im}\left(\alpha_i^{(0)}\right)}{\left|\alpha_i^{(0)}\right|^2}\frac{\mathrm{Im}\left(\alpha_i^{(\mathrm{d})}\right)}{\left|\alpha_i^{(\mathrm{d})}\right|^2} n(\omega,T_i)\mathrm{Tr}\left(\mathbb{A}_{ii}^{-1}\right)\right]\end{aligned} \tag{22}$$

At thermal equilibrium when all temperatures are equal, the net power absorbed must be zero for each frequency. This provides the condition [10]

$$\begin{aligned}0 &= \sum_j \frac{\left|\alpha_j^{(\mathrm{d})}\right|^2 \mathrm{Im}\left(\alpha_j^{(0)}\right)}{\left|\alpha_j^{(0)}\right|^2}\frac{\mathrm{Im}\left(\alpha_i^{(0)}\right)}{\left|\alpha_i^{(0)}\right|^2} \mathrm{Tr}\left(\mathbb{A}_{ij}^{-1}\mathbb{A}_{ij}^{-1\dagger}\right) \\ &- \frac{\left|\alpha_i^{(\mathrm{d})}\right|^2 \mathrm{Im}\left(\alpha_i^{(0)}\right)}{\left|\alpha_i^{(0)}\right|^2}\frac{\mathrm{Im}\left(\alpha_i^{(\mathrm{d})}\right)}{\left|\alpha_i^{(\mathrm{d})}\right|^2} \mathrm{Tr}\left(\mathbb{A}_{ii}^{-1}\right)\end{aligned} \tag{23}$$

Equation (23) enables us to arrive at a final expression for the net radiation heat transfer to the $i$th dipole in terms of the exchanges with all other dipoles:

$$Q_i(t,T_1,\ldots,T_N) = \int_0^\infty \frac{d\omega}{2\pi}\hbar\omega \sum_{j\neq i}\frac{4\left|\alpha_j^{(\mathrm{d})}\right|^2 \mathrm{Im}\left(\alpha_j^{(0)}\right)\mathrm{Im}\left(\alpha_i^{(0)}\right)}{\left|\alpha_j^{(0)}\right|^2\left|\alpha_i^{(0)}\right|^2}\mathrm{Tr}\left(\mathbb{A}_{ij}^{-1}\mathbb{A}_{ij}^{-1\dagger}\right) n_{ji}(\omega) \tag{24}$$

where $n_{ji}(\omega) = n(\omega,T_j) - n(\omega,T_i)$.

As expected, Equation (24) is very similar in form to the net heat transfer equation derived in the many-body theory [10] because we have followed their procedures for the latter part of the derivation. An advantage of Equation (24), however, is that it does not prescribe a treatment to correct for radiation damping since it is written in terms of the dressed polarizability. To demonstrate that our result is identical to that in reference [10], we can use the relation of the reduced absorption factor to the bare and dressed polarizabilities:

$$\chi_i = \frac{\left|\alpha_i^{(\mathrm{d})}\right|^2 \mathrm{Im}\left(\alpha_i^{(0)}\right)}{\left|\alpha_i^{(0)}\right|^2} \tag{25}$$

A proof of the equivalence of Equation (25) to the reduced absorption factor used in the many-body theory [10] is provided in Appendix C. Equation (25) allows us to rewrite Equation (24) as

$$Q_i(t,T_1,\ldots,T_N) = \int_0^\infty \frac{d\omega}{2\pi}\hbar\omega \sum_{j\neq i}\frac{4\chi_i\chi_j}{\left|\alpha_i^{(d)}\right|^2}\mathrm{Tr}\left(\mathbb{A}_{ij}^{-1}\mathbb{A}_{ij}^{-1\dagger}\right) n_{ji}(\omega) \tag{26}$$

When the dipoles are near thermal equilibrium, this may be written as the thermal conductance between two particles in a Landauer-like form [31]:

$$G_{ij}(T) = \int_0^\infty \frac{d\omega}{2\pi} \frac{\partial \Theta(\omega, T)}{\partial T} \mathcal{T}_{ij}(\omega) \tag{27}$$

where $\Theta(\omega, T_i) = \hbar\omega / n_i(\omega, T_i)$ is the mean energy of a harmonic oscillator and the transmission coefficient $\mathcal{T}_{ij}(\omega)$ is

$$\mathcal{T}_{ij}(\omega) = \frac{4\chi_i \chi_j}{\left|\alpha_i^{(d)}\right|^2} \mathrm{Tr}\left(\mathbb{A}_{ij}^{-1} \mathbb{A}_{ij}^{-1\dagger}\right) \tag{28}$$

We have obtained the same equation for many-dipole radiation as derived in the many-body method [10] while using the definitions of fluctuating and induced dipole moments used in the T-DDA [21], demonstrating their mathematical equivalence. Furthermore, we have done so with a straightforward use of the fluctuation-dissipation theorem without the need to derive the correlation function from the interactions of a dipole with a thermal bath. Although the choice of methodology may be a purely formal one, this derivation highlights some of the subtleties encountered when considering thermal-electromagnetic interactions, and it should clarify appropriate uses of the fluctuation-dissipation theorem in future research.

## III. Radiative Thermal Conductivity of Nanoparticle Arrays

### A. Radiative Thermal Conductivity Model

Although the previous equations provide energy absorption by a dipole or exchanges between pairs of dipoles, these quantities alone may not be particularly useful for very large systems containing many nanostructures, such as packed nanoparticle beds or nanofluids. In these cases, it is often desirable to compare the radiative contribution with the phonon contribution to the thermal transport [43]. When the systems are large enough that the spatial dimensions exceed modal propagation lengths, the thermal transport is diffusive and the most commonly used property is the thermal conductivity. We have previously derived expressions for the diffusive radiative thermal conductivity of linear chains of nanoparticles in order to compare the many-body method to a kinetic theory approach [47,48]. Here we generalize this method to arbitrary three-dimensional systems of dipoles in the diffusive limit.

We consider a collection of nanoparticles modeled as point dipoles at locations $\mathbf{r}_i$ with the same assumptions as discussed in Section II. An example of a portion of an ordered three-dimensional array is shown in Fig. 1, but the particles may also be disordered. We assume that all particles are near thermal equilibrium such that the thermal conductance between any two particles may be calculated with Equation (27). A small, linear temperature gradient is assumed to exist in the direction of interest for mathematical purposes, which is taken as the $x$ direction here. Note that if the particles are disordered or anisotropic, then different directions may exhibit different radiative thermal conductivities. In the center of the particle array, perpendicular to the temperature gradient, a fictitious plane is constructed to separate the array into two halves.

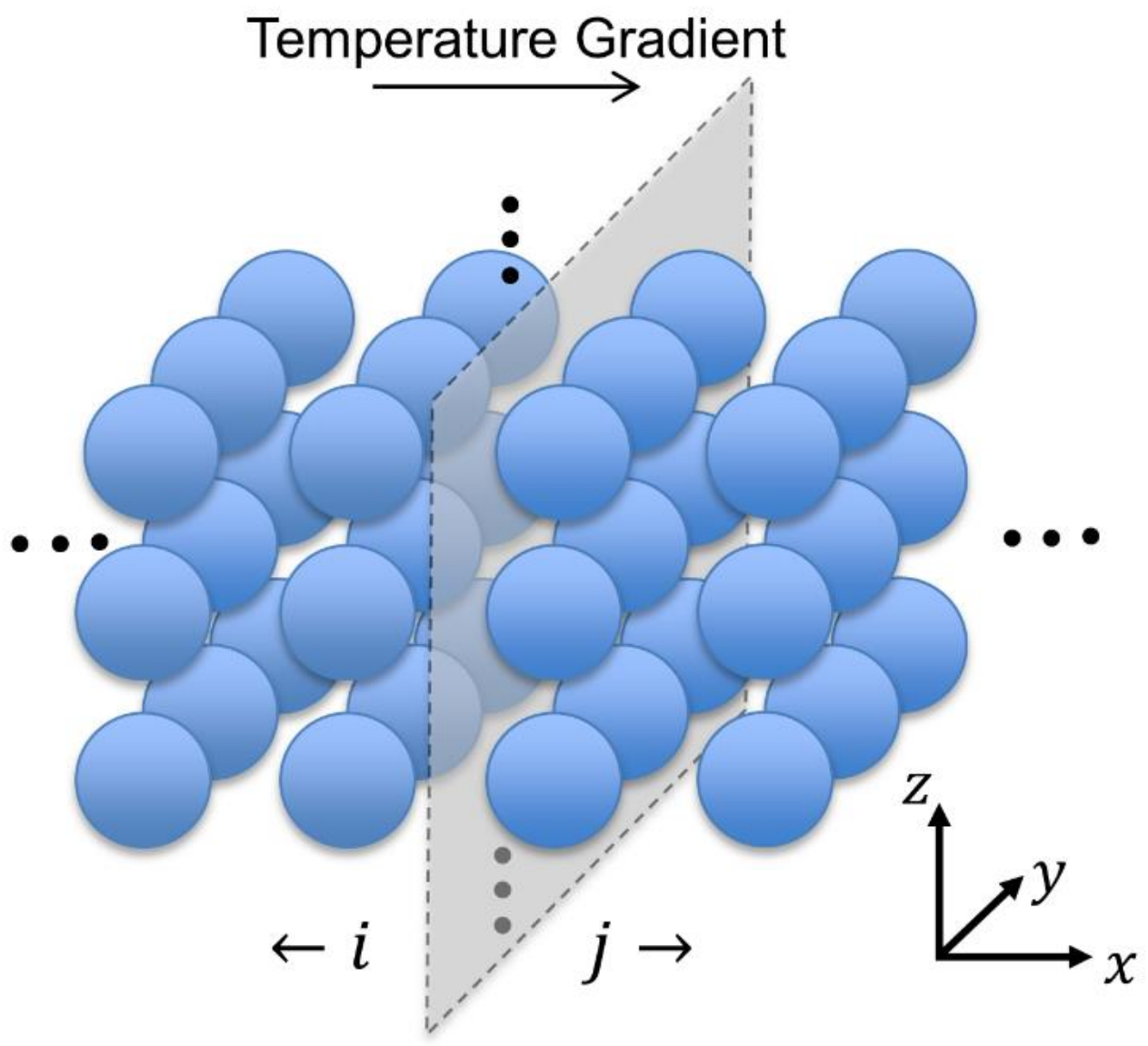

**Fig. 1.** Example of a portion of an ordered array of nanoparticles for the calculation of diffusive radiative thermal conductivity. A small, linear temperature gradient is assumed in the direction of interest, and a fictitious separating plane is introduced over which the radiative heat transfer is totaled. The total heat transfer is then converted into a thermal conductivity via Fourier's law.

In this scenario, the total heat transfer crossing the plane can be written in terms of the thermal conductance between all pairs of particles on opposing sides of the plane as

$$Q_{tot} = \sum_{i}\sum_{j} G_{ij}(T)\frac{dT}{dx}(\mathbf{r}_j - \mathbf{r}_i)\cdot\hat{x} \tag{29}$$

where the summation over $i$ runs over one side of the plane, the summation over $j$ runs over the other side, $T$ is the equilibrium temperature of all the particles, and $dT/dx$ is the temperature gradient. The $dT/dx\,(\mathbf{r}_j - \mathbf{r}_i)\cdot\hat{x}$ term provides a mathematical temperature difference which scales the thermal conductance appropriately with separation distance. Using Fourier's law $Q_{tot} = \kappa\, S\, dT/dx$, we obtain the effective radiative thermal conductivity:

$$\kappa(T) = \frac{1}{S}\sum_{i}\sum_{j} G_{ij}(T)(\mathbf{r}_j - \mathbf{r}_i)\cdot\hat{x} \tag{30}$$

where $S$ is the cross-sectional area of the array perpendicular to the temperature gradient. This equation for thermal conductivity has two important limitations. First, the assumption that all particles are near thermal equilibrium means that this approach is not valid for large temperature gradients through the array. Due to both the nonlinear nature of thermal radiation with temperature and superdiffusive effects that can exist for polaritonic particles [31], large temperature differences result in nonlinear temperature gradients. Second, the placement of the separating plane and direction of the temperature gradient may affect the results if there is significant spatial variation in particle arrangements or properties. For example, if the particles are closely packed on the left side of the array and become very sparse towards the right side, the arrangement is highly asymmetrical and results will be influenced by the location of the plane. For this reason, the approach is best suited to ordered arrays or disordered arrays with a consistent packing density.

Because we are interested in the bulk radiative thermal conductivity of a collection of particles, a sufficient number of particles must be included in the array to be representative of the larger system. Contributions to the radiative thermal conductivity from individual pairs of particles decrease as the distance between them increases. This allows one to begin with a small number of particles on each side of the plane and add additional particles while repeating the thermal conductivity calculation. As particles are added in all directions, the influence of boundary effects becomes smaller and the array becomes more representative of a larger system, causing the calculation to eventually converge. This process is simpler, of course, for an ordered array; for a disordered array, a representative particle arrangement must be used, and multiple representative arrangements should be checked.

One remaining important point is that for certain ordered arrays, symmetry may be exploited to reduce the computational burden. For example, if all particles are isotropic, consist of the same material, and are organized in an ordered cubic structure as illustrated in Figure 1, then the thermal conductance calculation in Equation (30) would be highly redundant. By recognizing repeated terms and lumping them with an appropriate multiplicative factor, we can substantially reduce the number of terms in the double summation.

## B. Materials and Geometry

We examine arrays of spherical SiC and $SiO_2$ nanoparticles of radius $a = 25$ nm in a transparent background medium of $\varepsilon_m = 4$ for SiC and $\varepsilon_m = 1$ for $SiO_2$. The background permittivity for SiC was chosen because this maximizes electromagnetic coupling as shown in reference [49]. Additionally, we have previously shown that under these conditions with a center-to-center spacing of $d = 3a$, SiC chains support propagating surface polaritons that dominate the heat transfer, while $SiO_2$ chains do not support these propagating surface modes [47,49]. This difference allows a comparison between the two systems to investigate the role of propagating modes in many-particle arrays. The particles are at temperature $T =$ 500 K. The relative permittivity of SiC is described with a Lorentz model [50]:

$$\varepsilon(\omega) = \varepsilon_\infty \left(1 + \frac{\omega_{LO}^2 - \omega_{TO}^2}{\omega_{TO}^2 - \omega^2 - i\omega\Gamma}\right) \tag{31}$$

where $\varepsilon_\infty = 6.7$, $\omega_{LO} = 1.82\times10^{14}$ rad s$^{-1}$, $\omega_{TO} = 1.49\times10^{14}$ rad s$^{-1}$, and $\Gamma = 8.92\times10^{11}$ rad s$^{-1}$. The relative permittivity of $SiO_2$ is taken from Palik's data [50] with piecewise cubic interpolation. The geometries examined are ordered one-, two-, and three-dimensional arrays of particles in a cubic lattice structure, as well as one-dimensional disordered chains. For the disordered chains, the particles are randomly perturbed in the $\hat{z}$ direction (assuming the ordered chain lies on the $x$ axis) by a distance on the interval $\left[-L_p, L_p\right]$ where $L_p$ is randomly selected as $0 \le L_p \le d$. This allows the generation of many chain arrangements with varying degrees of disorder. These geometries are illustrated in Fig. 2. In order to obtain results representative of a larger system, particles are added repeatedly to the arrays and the thermal conductivity is calculated until a 1% convergence criterion is reached.

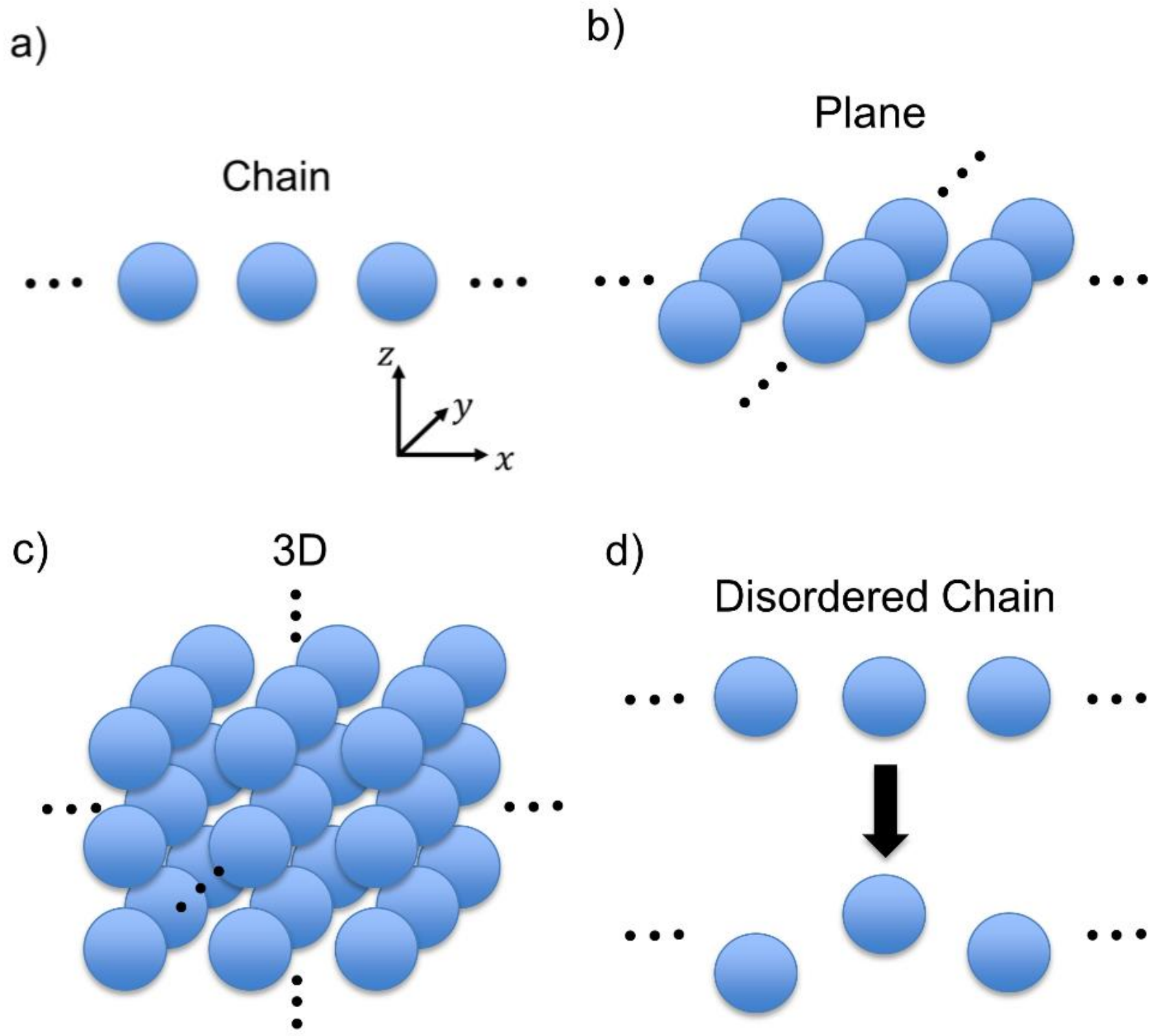

**Fig. 2.** Schematics of the geometries considered including (a) ordered one-dimensional chains, (b) ordered two-dimensional planes, (c) ordered three-dimensional arrays, and (d) disordered chains where the particles are perturbed in a direction perpendicular to the chain.

### C. Comparison to Exact Solution for Ordered Chains

To validate the many-dipole radiation methodology and to investigate its regime of validity, we compare the radiative thermal conductivity of ordered $SiO_2$ particle chains with varying $d$ to the exact solution derived by Czapla and Narayanaswamy [19]. Their solution is based on numerically exact vector spherical wave expansions of the dyadic Green's functions for particles in a linear chain. We use their code to calculate the thermal conductance between every pair of particles in the chain, and we transform these conductance values to a radiative thermal conductivity as described in Section III.A. For spacings $d < 3a$, a chain of 14 particles is used based on convergence at the minimum spacing of $d = 2a$. For spacings $d \geq 3a$, a chain of 10 particles is used based on convergence at a spacing of $d = 3a$. For these calculations, the area in the thermal conductivity calculation is taken as $S = \pi a^2$. The resulting radiative thermal conductivities are shown in Fig. 3.

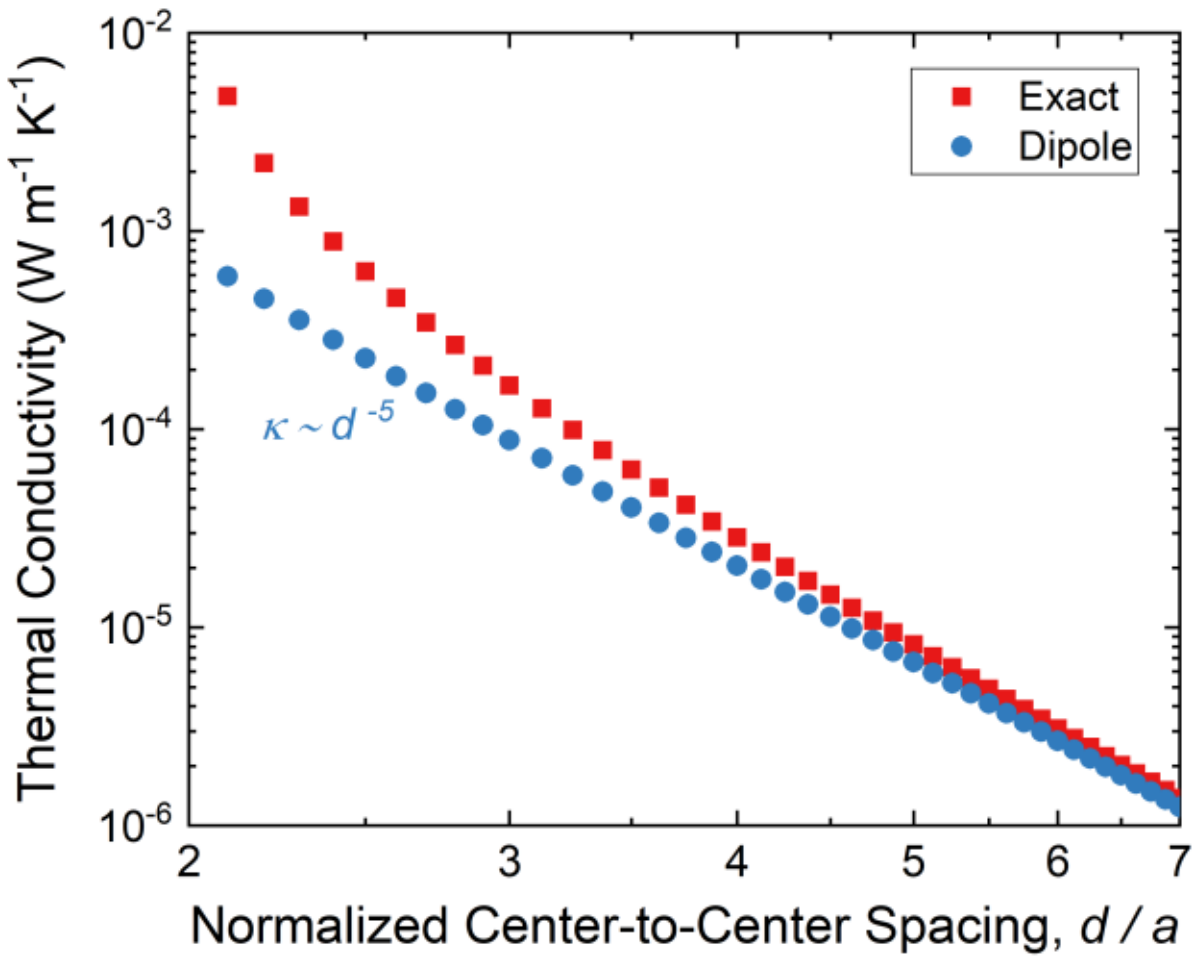

**Fig. 3.** Radiative thermal conductivity of ordered $SiO_2$ nanoparticle chains with varying center-to-center spacing calculated via the many-dipole method and the exact solution [19]. The dipole approximation underpredicts the exact results at small spacings and is commonly used down to spacings of $d \approx 3a$.

For the dipole method, we observe a typical monomial dependence on the spacing of $\kappa \sim d^{-n}$ with $n = 5$. This is consistent with the thermal conductance for dipole-dipole interactions [51] multiplied by distance when obtaining thermal conductivity. The dipole method and exact solution agree well for large separation distances, but the dipole method greatly underpredicts the exact solution at small spacings. This is a well-known effect, and typically researchers consider the dipole approximation to be valid for $d \gtrsim 3a$ [1,5,51,52]. For our purposes, the comparison of the exact and many-dipole methods provides two important results: it validates the theoretical formalism of the many-body radiative heat transfer theory, and it demonstrates that the dipole approximation may be used as a conservative baseline estimate of thermal radiation among closely spaced particles. The results from the exact method also show that near-field thermal radiation could be a significant contributor to thermal transport among very closely-spaced or packed particle arrays; for the case shown in Fig. 3, the exact radiative thermal conductivity is approaching values similar to those expected for phonon conduction between particles in contact [53].

## D. Ordered Particle Arrays

For a spacing of $d = 3a$, we calculate the spectral radiative thermal conductivity of $SiO_2$ and SiC particles in ordered chain, plane, and three-dimensional arrays as illustrated in Fig. 2(a), (b), and (c). The results are shown in Fig. 4. 10 particles for the $SiO_2$ chain and 22 particles for the SiC chain are used based on a 1% convergence criterion. These dimensions are extended to the plane and three-dimensional systems, resulting in $10^2 = 100$ ($22^2 = 484$) particles for the plane and $10^3 = 1{,}000$ ($22^3 = 10{,}648$) particles for the three-dimensional $SiO_2$ (SiC) cases. To fairly compare the three different geometries, we use cross-sectional areas for the radiative thermal conductivity in each calculation of $S = N_y N_z d^2$ where $N_y$ and $N_z$ are the lengths of the array in number of particles in the $y$ and $z$ directions. This results in lower thermal conductivities for the plane and chain than would be obtained considering the area as the physical extent of the system, but it allows the different cases to be compared directly to each other without differences due to the area.

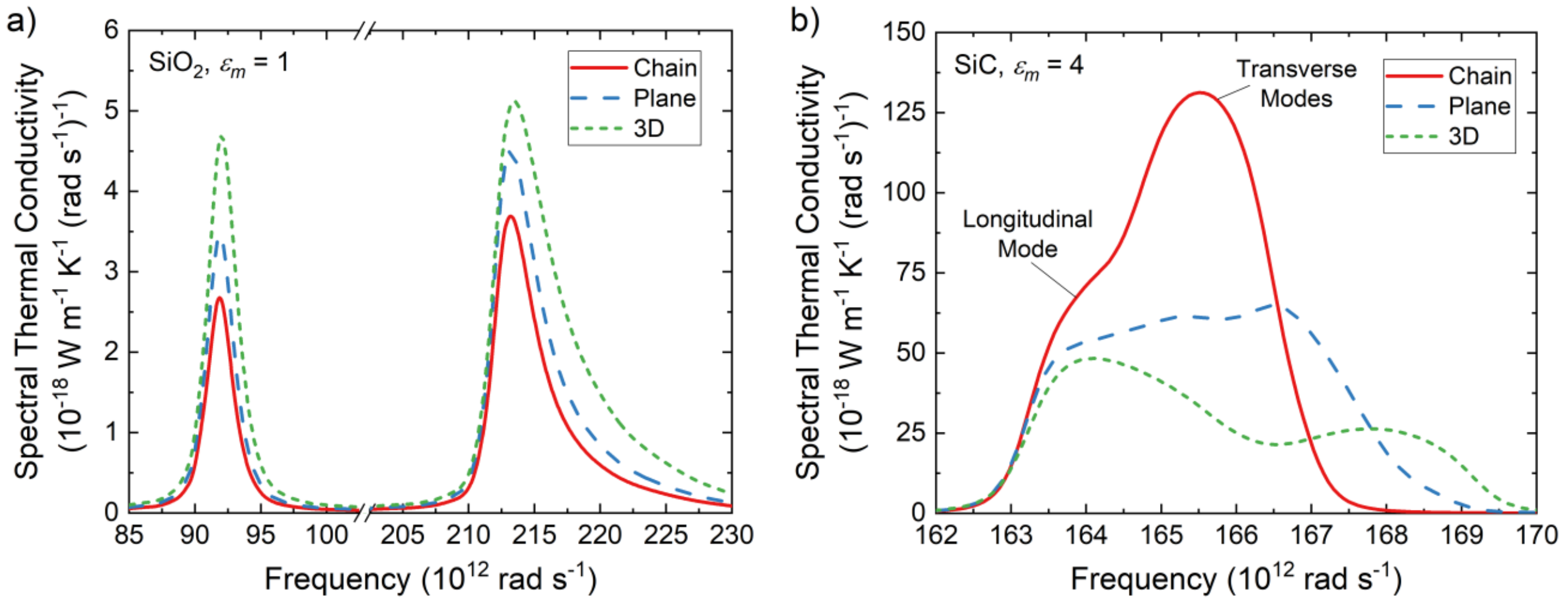

**Fig. 4.** Spectral radiative thermal conductivity of ordered chains, planes, and three-dimensional arrays of (a) $SiO_2$ particles in $\varepsilon_m = 1$ and (b) SiC particles in $\varepsilon_m = 4$. The spacing for all cases is $d = 3a$.

For the $SiO_2$ particles, we observe an increase in radiative thermal conductivity as dimensions are added to the particle chain. On the other hand, the SiC particles show a decrease in radiative thermal conductivity as dimensions are added to the array. These results reflect the predicted trends for thermal emission from ordered particle arrays [49]. The dissimilar behaviors are likely due to the existence or absence of propagating surface modes in the particle arrays. The $SiO_2$ particles do not support propagating modes [47,49], so the radiative thermal transport is due to the summation of particle-particle interactions in the array. Adding dimensions to the array increases the number of interactions and so increases the total thermal transport. The SiC particles do support propagating modes, which for a chain include a longitudinal mode corresponding to the shoulder in Fig. 4(b) and two degenerate transverse modes corresponding to the peak in Fig. 4(b) [47,49]. As dimensions are added to the SiC chain, transverse polarization can now couple to adjacent particles in that direction, creating a mode propagating perpendicular to the temperature gradient which is not a strong contributor to the radiative thermal conductivity. This may explain why the contribution of the longitudinal mode is relatively invariant while those of the transverse modes sharply decrease in Fig. 4(b) with increasing number of dimensions.

The results in Fig. 4 also have important implications for the general understanding of many-body effects in thermal radiation. Researchers have previously found that collective behavior tends to enhance thermal transport for systems containing a few particles [3,15] and to either enhance or suppress thermal transport for larger systems of particles depending on their number, positions, and optical properties [17]. Our results are consistent with the latter finding but provide physical insight into one mechanism by which many-body effects enhance or suppress thermal radiation. The ordered SiC arrays considered here enable coherence of the fluctuating thermal fields expressed through the collective behavior of propagating surface phonon polaritons, which is one reason why the spectral thermal conductivities are so much higher than for $SiO_2$. These modes can clearly be disrupted by controlling the placement of neighboring particles as shown in Fig. 4(b). Other methods of disrupting the propagating modes are by introducing disorder, which is examined in Section III.E, and by controlling the interparticle spacing, which is shown in Fig. 5. Here we plot the total radiative thermal conductivity of the $SiO_2$ and SiC systems as a function of the normalized center-to-center spacing $d/a$. For $d/a \geq 3$, we use the same numbers of particles as described for Fig. 4. For $d/a < 3$ we employ a 1% convergence criteria for a chain at $d = 2a$ to obtain 14 (62) particles in a chain, $14^2 = 196$ ($62^2 = 3{,}844$) particles in a chain, and $14^3 = 2{,}744$ ($62^3 = 238{,}328$) particles in a three-dimensional array for the $SiO_2$ (SiC) cases. It should be noted that the dipole approximation used here is expected to underpredict the thermal conductivity for $d/a < 3$ (indicated by the gray shaded region), as discussed in Section III.C. As before we use a cross-sectional area of $S = N_y N_z d^2$ for all cases for fair comparison across geometries.

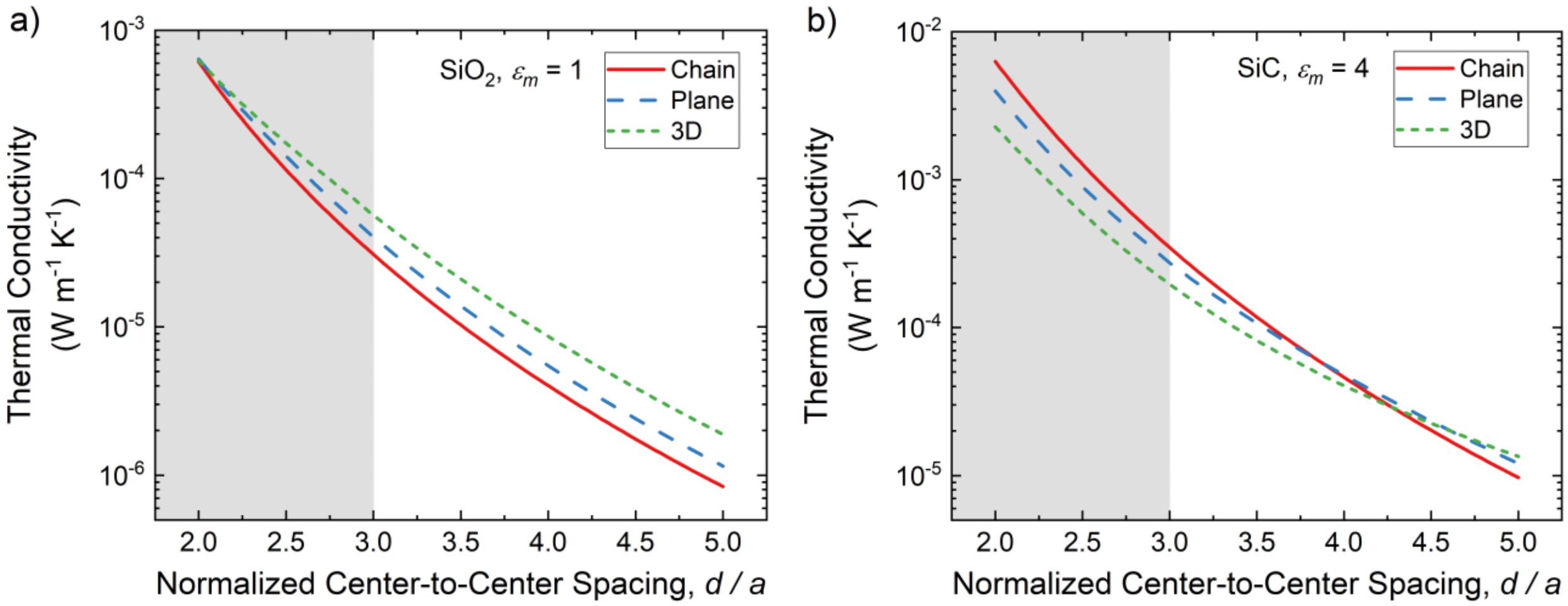

**Fig. 5.** Dependence of radiative thermal conductivity on spacing for ordered chains, planes, and three-dimensional arrays of (a) $SiO_2$ particles in $\varepsilon_m = 1$ and (b) SiC particles in $\varepsilon_m = 4$. Crossing points for the different geometries indicate a change between particle-particle interactions and collective interactions. The shaded gray region indicates where the dipole approximation substantially underpredicts the thermal conductivity.

At a spacing $d = 3a$, the total thermal conductivities reflect the differences between Fig. 4(a) and (b), with the highest $\kappa$ for $SiO_2$ being the three-dimensional case and the highest $\kappa$ for SiC being the chain case. As the spacing changes, however, the thermal conductivities for the different geometries become the same at about $d = 2a$ for $SiO_2$ and about $d = 4.25a$ for SiC. At larger spacings for SiC, the relative values of $\kappa$ switch for different geometries, with the highest being a three-dimensional array and the lowest being a chain. This change is likely due to a shift from collective, propagating modes being the dominant heat carriers to individual particle-particle exchanges being the dominant heat transfer mechanism. At larger spacing the particles cannot couple as strongly, reducing the contributions from propagating modes. Similarly for $SiO_2$, propagating modes are not supported at most spacings but begin to emerge as the particles are brought close to contact. These results reinforce the finding that many-body effects may enhance or suppress thermal radiation, but they are strongly dependent on the geometry of the particle array.

## E. Disordered Particle Arrays

Although only ordered particle arrays have been examined so far, in practice these are more difficult to assemble than disordered particle arrays. Disordered arrays are also more challenging to analyze, as they lack the symmetry of ordered arrays and additional steps are needed to determine realistic packing arrangements [54]. Nevertheless, we can determine how disorder tends to affect radiative thermal conductivity with a simplified case for a chain of particles. We consider chains of $SiO_2$ and SiC particles with spacing $d = 3a$, and disorder is introduced as illustrated in Fig. 2(d) and described in Section III.B. This spacing is selected for its reasonable accuracy with the dipole approximation. By only perturbing particles in a direction perpendicular to the axis chain, the separation distance between any two particles after randomization is $d \geq 3a$. The same numbers of particles are used as described for Fig. 4, and a cross-sectional area of $S = d^2$ is used again to permit fair comparison to the ordered chain. For both material systems, 1,000 random chain arrangements are generated, and the resulting thermal conductivities are shown in Fig. 6 as a fraction of the thermal conductivity of a perfectly ordered chain. We plot these results as a function of the weighted standard deviation of the particle displacements, where the particles closer to the dividing plane (as depicted in Fig. 1) are weighted more heavily than the particles further away. This type of weighting is used because the nearby particles have a much stronger influence on the calculated thermal conductivity, as shown by Equation (30). Because dipole contributions to the thermal conductivity tend to fall off as $d^{-5}$ as demonstrated in Fig. 3, the weighting function selected for the standard deviation is $[d/(d + |\mathbf{r}_j - \mathbf{r}_p|)]^{-5}$, where $\mathbf{r}_p$ is the location of the particle next to the dividing plane and $\mathbf{r}_j$ is the location of a particle further from the dividing plane. This function provides a weight of 1 for the particles adjacent to the plane and lower weights for particles further away.

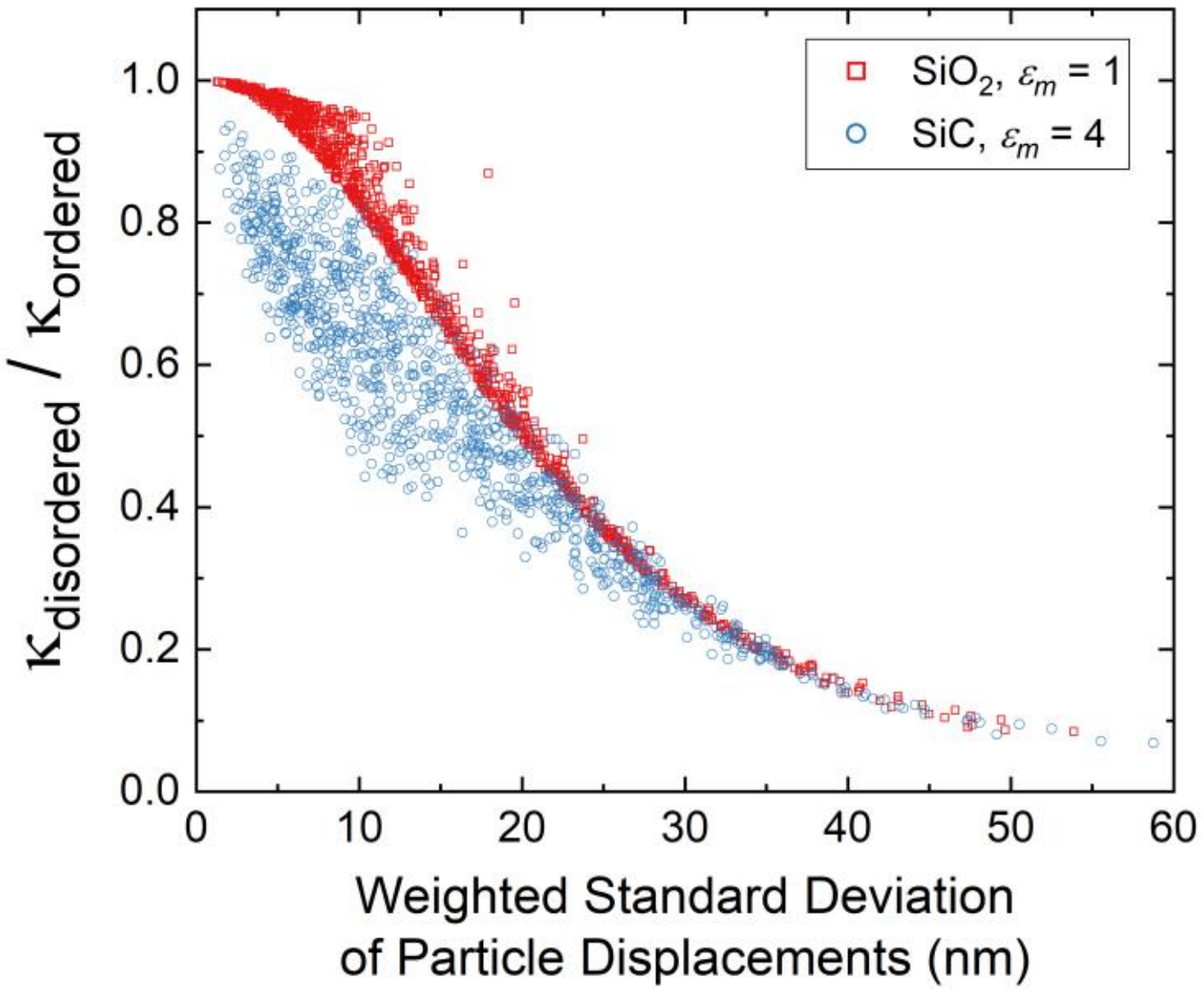

**Fig. 6.** Thermal conductivities of disordered $SiO_2$ and SiC particle chains as a fraction of the thermal conductivity of an ordered chain. The results are plotted as a function of the standard deviation of particle displacements, with particles closer to the dividing plane weighted more than particles further away. The steeper decrease seen for the SiC particles at low disorder indicates disruption of the propagating modes.

An important observation from the results is that in no case does the thermal conductivity of a disordered chain exceed that of an ordered chain. Although this is not surprising, it clearly demonstrates that ordered arrays are the best-case scenario when seeking to maximize radiative thermal transport. For both materials, dipole-dipole interactions become weaker as disorder increases because individual pairs of particles are further apart. For SiC, an additional effect is the disruption of propagating modes; when the SiC particles are no longer regularly spaced, coherence of the thermal fields is lost. This causes a more substantial decrease in the thermal conductivity at low levels of disorder when compared to $SiO_2$. The results shown here are likely to extend to disorder in two- and three-dimensional arrays of particles, especially for SiC or other materials that support propagating surface polaritons. For material systems like $SiO_2$ where particle pair interactions dominate the heat transfer, additional multipolar effects or electromagnetic screening effects may exist for disordered two- and three-dimensional arrays, which is a topic for future investigation.

## IV. Conclusions

Thermal radiation between nanostructures or discretized subvolumes is a topic of growing interest with a variety of applications. When these structures are modeled as point dipoles, two theoretical formalisms (the many-body method and the T-DDA) have emerged that use different forms of the fluctuation-dissipation theorem. We have shown that these two formalisms are mathematically equivalent. Furthermore, we have demonstrated that a straightforward use of the fluctuation-dissipation theorem is appropriate with the T-DDA approach, and the reduced absorption factor used with the many-body method is needed due to its definition of the induced and fluctuating dipole moments. These clarifications should assist future researchers in using correct forms of the fluctuation-dissipation theorem. To compare results from radiation calculations to other forms of heat transfer in large nanoparticle arrays, we have developed a method to calculate the effective radiative thermal conductivity from particle-particle thermal conductance. This method was used to analyze ordered chains, planes, and three-dimensional arrays of $SiO_2$ and SiC nanoparticles. Comparison of the two material systems demonstrated that the radiative transport is strongly influenced by whether the materials support propagating surface polariton modes. Additionally, we showed that many-body effects may enhance or suppress radiative transport depending on the geometry, spacing, and optical properties of the materials. Our results can help in the design of systems that utilize thermal radiation as a significant form of heat transfer. Finally, we demonstrated that ordered nanoparticle chains always exhibit higher radiative thermal conductivities than disordered particle chains due to disruption of propagating modes or increased distance between particles. Additional studies should focus on the impact of multipolar effects at small spacing and in disordered arrays, the effects of nonuniform size distribution in particle arrays [55], and methods to enhance thermal radiation in nanostructure arrays such as the use of nonhomogeneous environments [9].

## Acknowledgements

E.J.T. acknowledges support from the National Science Foundation Graduate Research Fellowship Program under Grant No. DGE-1650044. Any opinions, findings, and conclusions or recommendations expressed in this material are those of the authors and do not necessarily reflect the views of the National Science Foundation. Z.M.Z. would like to thank support from the U.S. Department of Energy, Office of Science, Basic Energy Sciences (DE-SC0018369). We would like to thank the Georgia Tech Partnership for an Advanced Computing Environment for their support with high performance computing.

## Appendix A: Dipole Moments in the T-DDA

For completeness we provide here a derivation of the induced part of the dipole moment shown in Equation (7) as well as the fluctuating part as used in the T-DDA. We begin with the differential forms of Faraday's law and Ampère's law in the $i$th particle or subvolume:

$$\begin{gathered}\nabla \times \mathbf{E}_i = i\omega\mu_0\mathbf{H}_i \\ \nabla \times \mathbf{H}_i = \mathbf{J}_i^{(\mathrm{fl})} + \mathbf{J}_i^{(\mathrm{ind})} - i\omega\varepsilon_0\varepsilon_i'\mathbf{E}_i\end{gathered} \tag{32}$$

where the currents $\mathbf{J}_i^{(\mathrm{fl})}$ and $\mathbf{J}_i^{(\mathrm{ind})}$ have been split into parts due to thermal fluctuations and due to the electric field, and $\varepsilon_i'$ is the real part of the relative permittivity. With the constitutive relation $\mathbf{J}_i^{(\mathrm{ind})} = \sigma\mathbf{E}_i$, where $\sigma$ is the electrical conductivity, and the definition of the complex relative permittivity $\varepsilon_i = \varepsilon_i' + i\frac{\sigma}{\omega\varepsilon_0}$, Ampère's law may be rewritten as

$$\nabla \times \mathbf{H}_i = \mathbf{J}_i^{(\mathrm{fl})} - i\omega\varepsilon_0\varepsilon_i\mathbf{E}_i \tag{33}$$

Taking the curl of Faraday's law and using Ampère's law, we obtain the vectorial wave equation:

$$\nabla \times \nabla \times \mathbf{E}_i - k_0^2\varepsilon_i\mathbf{E}_i = i\omega\mu_0\mathbf{J}_i^{(\mathrm{fl})} \tag{34}$$

where $k_0^2 = \omega^2\mu_0\varepsilon_0$. This may be written in terms of an equivalent source current as

$$\nabla \times \nabla \times \mathbf{E}_i - k_0^2\varepsilon_m\mathbf{E}_i = i\omega\mu_0\mathbf{J}_i^{(\mathrm{eq})} \tag{35}$$

where

$$\mathbf{J}_i^{(\mathrm{eq})} = -i\omega\varepsilon_0(\varepsilon_i - \varepsilon_m)\mathbf{E}_i + \mathbf{J}_i^{(\mathrm{fl})} \tag{36}$$

When the volume of the $i$th particle or subvolume is much smaller than the thermal wavelength, it may be modeled as a point dipole. The dipole moment is related to the current by [56]

$$\mathbf{p}_i = \frac{i}{\omega}\int_{V_i}\mathbf{J}_i dV \tag{37}$$

Inserting Equation (36) into Equation (37), we obtain the definition of the total dipole moment:

$$\mathbf{p}_i = \underbrace{V_i\varepsilon_0(\varepsilon_i - \varepsilon_m)\mathbf{E}_i}_{\mathbf{p}_i^{(\mathrm{ind})}} + \underbrace{\frac{i}{\omega}\int_{V_i}\mathbf{J}_i^{(\mathrm{fl})}dV}_{\mathbf{p}_i^{(\mathrm{fl})}} \tag{38}$$

In Equation (38), we have the induced and fluctuating parts of the dipole moment, and we see the bare polarizability $\alpha_i^{(0)} = V_i(\varepsilon_i - \varepsilon_m)$ appear in the induced part as shown in Equation (7).

## Appendix B: Fluctuation-dissipation Theorem from Thermal Currents

Equation (20) provides the fluctuation-dissipation theorem in terms of the particle bare polarizability as derived in reference [41] and summarized in reference [40]. This equation may also be derived directly from the fluctuation-dissipation theorem for the thermal currents as done in reference [21]. Beginning with the fluctuation-dissipation theorem [37-39] for the convention of the Fourier transform chosen, we have

$$\langle J_{j,\gamma}^{(\mathrm{fl})}(\omega)J_{j',\gamma'}^{(\mathrm{fl})*}(\omega')\rangle = 4\pi\hbar\omega^2\varepsilon_0 n(\omega, T_j)\mathrm{Im}(\varepsilon_j)\delta_{jj'}\delta_{\gamma\gamma'}\delta(\omega - \omega') \tag{39}$$

Assuming a small enough volume such that the current may be approximated as constant throughout, we may use Equation (37) to transform this into an equation for fluctuating dipole moments:

$$\langle p_{j,\gamma}^{(\mathrm{fl})}(\omega)p_{j',\gamma'}^{(\mathrm{fl})*}(\omega')\rangle = 4\pi\hbar V_j\varepsilon_0 n(\omega, T_j)\mathrm{Im}(\varepsilon_j)\delta_{jj'}\delta_{\gamma\gamma'}\delta(\omega - \omega') \tag{40}$$

Equation (40) is identical to Equation (20), because $V_j\varepsilon_0\mathrm{Im}(\varepsilon_j) = \mathrm{Im}(\alpha_j^{(0)})$.

## Appendix C: Reduced Absorption Factor

To transform our result for the net radiation heat transfer given by Equation (24) to the same form as that obtained in the many-body theory, we used a definition of the reduced absorption factor in terms of the dressed and bare polarizabilities in Equation (25) which is repeated here:

$$\chi_i = \frac{\left|\alpha_i^{(\mathrm{d})}\right|^2 \mathrm{Im}\left(\alpha_i^{(0)}\right)}{\left|\alpha_i^{(0)}\right|^2} \tag{41}$$

We prove that this definition is equivalent to that used in the many-body theory [10]:

$$\chi_i = \mathrm{Im}\left(\alpha_j^{(\mathrm{d})}\right) - \frac{k^3}{6\pi\varepsilon_m}\left|\alpha_j^{(\mathrm{d})}\right|^2 \tag{42}$$

We can write Equation (42) in terms of the dyadic Green's function $\mathbb{G}_{ii}^{(0)}$ as

$$\chi_i = \mathrm{Im}\left(\alpha_j^{(\mathrm{d})} - \frac{k^2}{\varepsilon_m}\mathbb{G}_{ii}^{(0)}\left|\alpha_j^{(\mathrm{d})}\right|^2\right) \tag{43}$$

Using the definition of the dressed polarizability from Equation (12), this becomes

$$\chi_i = \mathrm{Im}\left[\alpha_j^{(\mathrm{d})} + \left(\frac{1}{\alpha_j^{(\mathrm{d})}} - \frac{1}{\alpha_j^{(0)}}\right)\left|\alpha_j^{(\mathrm{d})}\right|^2\right] \tag{44}$$

With some algebraic manipulation and use of the identities $zz^* = |z|^2$ and $\mathrm{Im}(z^*) = -\mathrm{Im}(z)$, Equation (44) can be shown to be identical to Equation (41).